\documentclass[aps,prx,reprint,twocolum,amsmath,superscriptaddress,amssymb,showpacs,nofootinbib]{revtex4-2}

\usepackage[utf8]{inputenc}
\usepackage[english]{babel}
\usepackage{booktabs} 
\usepackage{comment} 
\usepackage{hyperref} 
\usepackage{physics} 
\usepackage[table,xcdraw,dvipsnames]{xcolor} 

\allowdisplaybreaks

\usepackage{lipsum}
\usepackage{xfrac}

 \usepackage{multirow}

\usepackage{amsmath,verbatim,latexsym,amssymb,indentfirst,mathrsfs,mathtools,amsthm,bbm,bm,hyperref,url,cancel,subcaption}
\usepackage[font=small,labelfont=bf,format=plain,justification=raggedright,singlelinecheck=false]{caption}
\usepackage{verbatim,indentfirst}
\usepackage[title,titletoc]{appendix}

\usepackage{graphicx}
\graphicspath{{images/}}
\usepackage{epstopdf}

\makeatletter
\def\p@subsection{}
\makeatother
\makeatletter
\def\p@subsubsection{}
\makeatother

\makeatletter
\newcommand\footnoteref[1]{\protected@xdef\@thefnmark{\ref{#1}}\@footnotemark}
\makeatother

\usepackage{soul}

\usepackage{color}
\usepackage[dvipsnames]{xcolor}
\usepackage[normalem]{ulem}

\makeatletter
\DeclareFontFamily{OMX}{MnSymbolE}{}
\DeclareSymbolFont{MnLargeSymbols}{OMX}{MnSymbolE}{m}{n}
\SetSymbolFont{MnLargeSymbols}{bold}{OMX}{MnSymbolE}{b}{n}
\DeclareFontShape{OMX}{MnSymbolE}{m}{n}{
    <-6>  MnSymbolE5
   <6-7>  MnSymbolE6
   <7-8>  MnSymbolE7
   <8-9>  MnSymbolE8
   <9-10> MnSymbolE9
  <10-12> MnSymbolE10
  <12->   MnSymbolE12
}{}
\DeclareFontShape{OMX}{MnSymbolE}{b}{n}{
    <-6>  MnSymbolE-Bold5
   <6-7>  MnSymbolE-Bold6
   <7-8>  MnSymbolE-Bold7
   <8-9>  MnSymbolE-Bold8
   <9-10> MnSymbolE-Bold9
  <10-12> MnSymbolE-Bold10
  <12->   MnSymbolE-Bold12
}{}

\let\llangle\@undefined
\let\rrangle\@undefined
\DeclareMathDelimiter{\llangle}{\mathopen}%
                     {MnLargeSymbols}{'164}{MnLargeSymbols}{'164}
\DeclareMathDelimiter{\rrangle}{\mathclose}%
                     {MnLargeSymbols}{'171}{MnLargeSymbols}{'171}
\makeatother
\usepackage{amsmath}
\usepackage[normalem]{ulem}
\usepackage{afterpage}
\usepackage{xcolor}[dvipsnames]

\begin{document}
\title{A unification of the coding theory and OAQEC perspectives on hybrid codes}
\author{Shayan Majidy} 
\email{smajidy@uwaterloo.ca}
\affiliation{Perimeter Institute for Theoretical Physics, Waterloo, Ontario N2L 2Y5, Canada} 
\affiliation{Institute for Quantum Computing and Department of Physics and Astronomy, University of Waterloo, Waterloo, Ontario, Canada}

\begin{abstract}
There is an advantage in simultaneously transmitting both classical and quantum information over a quantum channel compared to sending independent transmissions. The successful implementation of simultaneous transmissions of quantum and classical information will require the development of hybrid quantum-classical error-correcting codes, known as hybrid codes. The characterization of hybrid codes has been performed from a coding theory perspective and an operator algebra quantum error correction (OAQEC) perspective. First, we demonstrate that these two perspectives are equivalent and that the coding theory characterization is a specific case of the OAQEC model. Second, we include a generalization of the quantum Hamming bound for hybrid error-correcting codes. We discover a necessary condition for developing non-trivial hybrid codes---they must be degenerate. Finally, we construct an example of a non-trivial degenerate 4-qubit hybrid code.
\end{abstract}

\maketitle

\section{Introduction}


The simultaneous transmission of quantum and classical information over a quantum channel was initially explored in Ref.~\citep{devetak2005capacity}. There it was shown that there exists an advantage in transmitting both quantum and classical information simultaneously over a quantum channel when compared to independent transmissions. This work has since been followed by many authors \citep{hsieh2010entanglement, hsieh2010trading, kremsky2008classical, nemec2022hybrid, nemec2018hybrid, nemec2020nonbinary, nemec2021infinite}. Different error-correcting codes for the simultaneous transmission of quantum and classical information (``hybrid codes'') have been developed. There exist two characterizations of hybrid codes in the literature. One was developed from a coding theory perspective \citep{grassl2017codes} and the other from an operator algebra quantum error correction (OAQEC) perspective \citep{beny2007quantum}.

A simple example can illustrate the basic structure of a hybrid code. Consider a two-qubit system. In the computational basis, these two qubits will have four orthogonal states available to them; $\ket{00}$, $\ket{01}$, $\ket{10}$ and $\ket{11}$. From these states, we can construct two pairs of logical states, and each pair can be labelled with the classical bits $0$ or $1$:
\begin{align}
\ket{\bar{0}}_0 \coloneqq \ket{00}, \, \ket{\bar{1}}_0 \coloneqq\ket{01}, \,
\ket{\bar{0}}_1 \coloneqq \ket{10}, \, \ket{\bar{1}}_1 \coloneqq \ket{11}.
\end{align}
These four codewords can move through a two-qubit channel and be immune to $ZI$ errors. In this simple example, two unprotected qubits are used to transmit one classical bit and one qubit of information with some degree of error protection. The different hybrid codes which have been proposed (see \citep{grassl2017codes}  for examples) go well beyond this simple construction. The main text explains when hybrid code constructions provide an advantage over independent quantum and classical information transmissions. 
The objective of a quantum error correcting code is to (i) maximize the number of logical qubits which can be transmitted, (ii) maximize the robustness of their protection, and (iii) minimize the number of physical qubits that are required to transmit them. Hybrid codes add an additional dimension---maximizing the amount of classical information which is being transmitted.

We first demonstrate that the two perspectives on hybrid codes are equivalent and that the coding theory characterization is a specific case of the OAQEC model. This allows for each perspective to benefit from the results developed in the other. Next, we study the quantum Hamming bound. This bound is a seminal result in quantum error correction and loadstone for developing quantum error-correcting codes~\cite{gottesman1996class}. Extending such bounds to hybrid codes is thus an important step in developing hybrid codes. We generalize the quantum Hamming bound for hybrid error-correcting codes. Third, we discover a necessary condition for developing non-trivial hybrid codes---non-trivial hybrid codes must be degenerate. This is a clear restriction that all future non-trivial hybrid codes must adhere to. Finally, proof-of-principle experimental tests of hybrid codes have not been performed. One barrier to doing so is that the non-trivial codes to date require more qubits than are available in most academic labs~\cite{grassl2017codes}. We present an example of a non-trivial 4-qubit hybrid code.

The rest of this paper is organized as follows. Section \ref{sec:twopersp} reviews the two perspectives on hybrid codes, and Sec.~\ref{sec:unified} presents their unification. We present a generalization of the Hamming bound on quantum error correcting codes for hybrid codes in Sec.~\ref{sec:Hamming}. We demonstrate here that non-trivial hybrid codes must be degenerate and exemplify a 4-qubit example of such a code.

\section{Two perspectives} \label{sec:twopersp}

\subsection{Coding theory perspective}

The characterization and construction of hybrid codes were formulated from a coding theory perspective in Ref.~\citep{grassl2017codes}. A quantum error-correcting code that encodes $k$ qubits into $n$ qubits with a distance $d$ can be denoted by $C = [[n,k,d]]$ or equivalently as $C = ((n, K, d))$ where $K$ is the dimensions of the subspace of the $k$ qubits, $K = 2^k$. A classical code can be denoted identically, except with $m$ and $M$ representing the number of classical bits and dimension of the bits subspace respectively, $C = [[n,m,d]]$ or  $C = ((n, M, d))$. In this notation, a hybrid code can be denoted as $C = [[n,k:m,d]]$ or $C = ((n,K:M,d))$.

Using this notation, we can describe three  hybrid code constructions which do not provide an advantage over the independent solutions. First,  given a quantum code, $C = ((n, KM ,d))$, one can factor the code space into two subsystems of dimension $K$ and $M$, and use these subspaces to transmit $K$ dimensional quantum and $M$ dimensional classical information separately. Since this effectively sacrifices quantum bits for classical bits, there is no advantage to using such codes. A second construction comes from assuming one already has a hybrid code $C = [[n,k:m,d]]$ and constructing a new code $C' = [[n, k-1:m+1, d]]$. A qubit can always be used to transmit classical information, making this construction also trivial. Finally, given a quantum code $C_q = [[n_1, k, d]]$ and classical code $C_c = [[n_2, m, d]] $ one can form a hybrid code $C = [[n_1 + n_2, k:m, d]]$, which again would not provide any advantage. Our goal in developing hybrid codes is to find codes with better parameters than those provided.

A hybrid quantum code $C = ((n, K:M ,d))$ can be described by a collection of $M$ quantum codes $\{C^{(\nu)} : \nu = 1,..., M\}$, where $\nu$ is the classical information that determines which of the $C^{(\nu)}$ is used. Each code has an orthonormal basis $\{\ket{c_i^{(\nu)}}: i = 1,2,\hdots ,K\}$. Denote by $\{E_A\}$ the set of errors each code can correct. To correct such errors, each code must obey a modified version of the Knill--Laflamme condition \citep{knill1997theory}:
\begin{equation}
\mel{c_i^{(\nu)}}{E_{k}^{\dagger}E_{l}}{c_j^{(\nu)}} = \alpha_{kl}^{(\nu)}\delta_{ij} \label{eqn:KLcondh},
\end{equation}
where $\alpha_{kl}^{(\nu)}$ is a complex constant. Equation \eqref{eqn:KLcondh} differs from the original Knill--Laflamme condition in that the constant $\alpha_{kl}^{(\nu)}$ depends on the classical information being transmitted as well. A second condition hybrid codes must satisfy is that each quantum code must be simultaneously distinguishable from all others. This is necessary for retrieving the classical information. This provides a second error correction condition on hybrid codes
\begin{equation}
    \bra{c_i^{(\nu)}}E_{k}^{\dagger}E_{l}\ket{c_j^{(\mu)}} = 0, \quad \text{for} \quad \mu \neq \nu \label{eqn:Hdistinguish}
\end{equation}
Equations \eqref{eqn:KLcondh} and \eqref{eqn:Hdistinguish} can be written succinctly as
\begin{equation}
    \bra{c_i^{(\nu)}}E_{k}^{\dagger}E_{l}\ket{c_j^{(\mu)}} = \alpha_{kl}^{(\nu)}\delta_{ij}\delta_{\nu \mu} \label{eqn:Hcodecondition}
\end{equation}
The proof for this condition is outlined in Ref.~\citep{grassl2017codes}. 

\subsection{OAQEC perspective}
To introduce OAQEC, we must first introduce its predecessor---operator quantum error correction (OQEC) \citep{kribs2005unified}. OQEC unifies the standard model and the noiseless subsystems model of error correction. The standard model \citep{knill1997theory, shor1995scheme, steane1996simple, bennett1996ch} consists of the 3-tuple $(\mathcal{R}, \mathcal{E}, \mathcal{C})$. Consider a quantum system whose states are elements of the Hilbert space $\mathcal{H}$ and $\mathcal{B}(\mathcal{H})$ are the set of operators acting on $\mathcal{H}$. Denote by $\mathcal{C}$ the quantum code, a subspace of $\mathcal{H}$. $\mathcal{E}$ and $\mathcal{R}$ are the error and recovery operations acting on $\mathcal{B}(\mathcal{H})$, such that 
$\mathcal{R}$ corrects $\mathcal{E}$,
\begin{equation}
    (\mathcal{R} \circ \mathcal{E})(\rho) = \rho, \quad \forall \rho \in \mathcal{C}.
\end{equation}
Using the Kraus operator notation, we can write $\mathcal{E}(\rho) = \sum_a E_a \rho E_a^{\dagger}$. Denote by $P_{\mathcal{C}}$ the projection operator onto $\mathcal{C}$. Such an $\mathcal{R}$ exists for a given $\mathcal{E}$ and $\mathcal{C}$ if
\begin{equation}
    P_{\mathcal{C}}E_a^{\dagger}E_bP_{\mathcal{C}} = \lambda_{ab}P_{\mathcal{C}}, \quad \forall a,b,
\end{equation}
where $\lambda_{ab}$ is a constant~\cite{bennett1996ch, knill1997theory}.

In the noiseless subsystem model~\citep{palma1996quantum, duan1997preserving, zanardi1997noiseless, lidar1998decoherence}, one considers the error set they wish to correct $\{E_a\}$. Denote by $\mathcal{A}$ the algebra generated by $\{E_a,E_a^{\dagger}\}$. Since $\mathcal{A}$ is a C$^{*}$-algebra, it is unitarily equivalent to a direct sum of full matrix algebras, $\mathcal{A} = \bigoplus_J M_{m_j}\otimes I_{n_j}$. This decomposes the Hilbert space into a noisy and noiseless subsystem, which we denote as subsystems $A$ and $B$, respectively. Elements in the noiseless subsystem (the ``noise commutant'') will commute with $\mathcal{A}$ and be immune to errors. The information one wishes to protect is encoded in the noise commutant $\mathcal{A'} = \bigoplus_J I_{m_j}\otimes M_{n_j}$. By defining the projection operators $P_{kl} = \dyad{\alpha_k}{\alpha_l} \otimes I_n$, one can define a map into this noiseless subspace with the following properties,
\begin{align}
    \Gamma(\rho) &= \sum_{k,l} P_{kl} \rho P^{\dagger}_{kl} \in \mathcal{A}',\\
\Gamma(\rho_A \otimes \rho_B) &\propto I_A \otimes \rho_B.
\end{align}
This can be further generalized since it is not necessary to protect the entire space $\rho_A \otimes \rho_B$, only $\rho_B$. Thus instead of being confined to  $\mathcal{A'} = I_{A}\otimes \rho_B $ we consider the space $\mathcal{U} = \rho_A \otimes \rho_B$. We define $P_k \coloneqq P_{kk}$, $P_{\mathcal{U}} \coloneqq \sum_k P_k$, and $P^{\perp}_{\mathcal{U}} = I - P_{\mathcal{U}}$. With these, we can give three equivalent definitions for a noiseless subsystem:
\begin{align}
\forall \rho_A, \rho_B, \exists \sigma_A &: \mathcal{E}(\rho_A \otimes \rho_B) = \sigma_A \otimes \rho_B,\\
\forall \rho_B \exists \sigma_A &: \mathcal{E}(I_A \otimes \rho_B) = \sigma_A \otimes \rho_B, \, \text{and}\\
\forall \rho \in \mathcal{U} &: (\text{tr}_A \circ \mathcal{P}_\mathcal{U} \circ \mathcal{E})(\rho) = \text{tr}_A(\rho).
\end{align}
The subspace $\mathcal{H}_B$ is noiseless if it satisfies any, and thus all, of the above conditions. There exists such a semigroup $\mathcal{U}$ for a channel $\mathcal{E}$ if and only if, 
\begin{align}
P_kE_aP_l = \lambda_{akl}P_{kl}, \quad &\forall a,k,l, \, \text{and}\\
P_{\mathcal{U}}^{\perp}E_aP_{\mathcal{U}} = 0, \quad &\forall a.
\end{align}
OQEC also consists of the 3-tuple $(\mathcal{R}, \mathcal{E}, \mathcal{C})$. The noiseless subspace model is a specific case where $\mathcal{R} = I$, and the standard model is a specific case where  $\mathcal{C} = \mathcal{U}$. In OQEC, such a 3-tuple is correctable if
\begin{equation}
(\text{tr}_A \circ \mathcal{P}_\mathcal{U} \circ \mathcal{R} \circ \mathcal{E})(\rho) = \text{tr}_A(\rho)
\end{equation}
A necessary condition for the existence of such a $\mathcal{U}$ is
\begin{equation}
P_kE_i^{\dagger}E_jP_l = \alpha_{ijkl}P_{kl} \quad \forall i,j,k,l.
\label{eqn:OAQECecc}
\end{equation}

We now move from OQEC to OAQEC. In the Schr\"odinger picture, our states change with time. Thus the expectation value of an operator, $A$, acting on a system, $\rho$, which is experiencing an error,  $E_i$, is given by $\text{tr}({A E_i \rho E_i^{\dagger}}) = \text{tr}({ E_i^{\dagger} A E_i \rho})$. The right-hand side of the equation implies that the error is acting on the observable (a Heisenberg picture of the error). For every trace-preserving channel in the Schr\"odinger picture that acts on state $\rho$, there exists a corresponding dual map which is unital acting on the observable $A$. 

In the Schrodinger picture a subspace was noiseless for a channel $\mathcal{E}$ if $\mathcal{E}(\rho_A \otimes \rho_B) = \sigma_A \otimes \rho_B$, equivalently a space is noiseless for an error channel $\mathcal{E}^{\dagger}$ in the Heisenberg picture if and only if $P\mathcal{E}^{\dagger}(X \otimes I) P= X \otimes I$ for all $X$ which are observables. Where $P$ is the projector of the Hilbert space onto the subspace $A \otimes B$.  This gives two equivalent definitions for a noiseless subsystem; when one is satisfied, the other is also. 

We say a set of operators $S$ on $\mathcal{H}$ are conserved by $\mathcal{E}$ for states on some subspace $\mathcal{H}_S$ if every $X \in S$ satisfies $P\mathcal{E}^{\dagger}(X)P = PX P$. These observables can generate an algebra that we wish to protect from errors. This can equivalently be done via a theorem from  \citep{beny2007quantum}, which states: Let $\mathcal{A}$ be a subalgebra of $\mathcal{B}(\mathcal{H}_s)$. $\mathcal{A}$ is conserved by $\mathcal{E}$ if and only if $E_aP$ commutes with every element of the algebra. This can be expanded to the subalgebra $\mathcal{A}$ being correctable for $\mathcal{E}$ if and only if $PE_a^{\dagger}E_bP$ commutes with every element of the algebra for every combination of errors. This generalization, when now considered from the perspective of the Schr\"odinger picture, shows that the algebra $\mathcal{A}$ is correctable for $\mathcal{E}$ for subspaces of the Hilbert space $\mathcal{H}_S$ on one condition. The condition is that there exists a recovery operation $\mathcal{R}$ such that, for any density operator which can be separated into a sum of tensor products of operators in the separate spaces $\rho = \sum_k \alpha_k (\rho_k \otimes \tau_k)$ for $\sum_k \alpha_k = 1$, the following equation holds
\begin{equation}
    (\mathcal{R} \circ \mathcal{E})(\rho) = \sum_k \alpha_k \mathcal{R}(\mathcal{E}(\rho_k \otimes \tau_k)) = \sum_k \alpha_k (\rho_k \otimes \tau_k').
\end{equation}
Not that for $\alpha_1 = 1$, this reduces to the OQEC condition. $(\mathcal{R} \circ \mathcal{E})(\rho_k \otimes \tau_k) = \rho_k \otimes \tau_k'$. Denote by $X_{abk}$ operators acting on subspace $B$. There exists such a correction operation if and only if for all $a,b$ there exists $X_{abk}$ such that:
\begin{equation}
P E_a^{\dagger}E_b P = \sum_k I_{A_k} \otimes X_{abk}.
\end{equation}

\section{Unified perspective} \label{sec:unified}
We now show that the coding theory perspective is a special case of the OAQEC formulation. The coding theory construction included three restrictions that do not exist in the OAQEC model. First, the error set in the coding theory construction is restricted to containing only unitary errors, particularly the Pauli channel. Second, in the coding theory model each quantum code is viewed as a subspace where the OAQEC model deals with subsystems. Finally, the coding theory model restricts each quantum channel to be of equal dimension. These three restrictions can be summarized as:
\begin{enumerate}
  \item $\mathcal{E} \subseteq P_n$
  \item $C^{(\nu)} \subseteq \mathcal{H}, \forall \quad 1\leq \nu \leq M$ subspaces
  \item dim $C^{(\nu)} = K$ $\forall$ $\nu$, $C^{(\nu)} = \text{span} \{ \ket{c_i^{(\nu)}}: 1 \leq i \leq K \}$ 
\end{enumerate}
The coding theory error correction condition acts on codewords in Hilbert space while the OAQEC models acts on operators. In unifying these two models it is necessary to rewrite the coding theory condition in terms of operators. This can be done by considering the two equivalent forms of the regular Knill--Laflamme condition. 
\begin{center}
$\bra{c_i}E_a^{\dagger}E_b\ket{c_j} = \alpha_{ab}\delta_{ij}$ $\iff$ $P E_a^{\dagger}E_b P = \alpha_{ab}P$.
\end{center}
Either form can be used. With this, we can define the projector onto the hybrid codeword space as:
\begin{equation}
    P = \sum_{i ,\nu} \dyad{c_i^{(\nu)}}{c_i^{(\nu)}} \label{eqn:projectorsforBei}
\end{equation}
This projector can then be used to rewrite Eq.~\eqref{eqn:Hcodecondition} as
\begin{equation}
    P E_a^{\dagger}E_b P = \sum_{i,\nu}  \alpha_{kl}^{(\nu)} \dyad{c_i^{(\nu)}}{c_i^{(\nu)}} \label{eqn:rewrittenBei}
\end{equation}
Unlike the general Knill--Laflamme condition, $\alpha$ depends on the codewords, so it must be included within the projector on the right-hand side of Eq.~\eqref{eqn:projectorsforBei}.  Consider substituting Eq.~\eqref{eqn:projectorsforBei} into Eq.~\eqref{eqn:rewrittenBei}:
\begin{align}
    P E_a^{\dagger}E_b P &= \sum_{i,j,\nu,\mu} \dyad{c_i^{(\nu)}}{c_i^{(\nu)}}E_a^{\dagger}E_b \dyad{c_j^{(\mu)}}{c_j^{(\mu)}},\\
    P E_a^{\dagger}E_b P &= \sum_{i,j,\nu,\mu}  \alpha_{ab}^{(\nu)} \delta_{ij} \delta_{\nu \mu} \dyad{c_i^{(\nu)}}{c_j^{(\mu)}},\\
    P E_a^{\dagger}E_b P &= \sum_{i,\nu} \alpha_{ab}^{(\nu)} \dyad{c_i^{(\nu)}} {c_i^{(\nu)}}.\label{eqn:hybridrewritten}
\end{align}
By starting from the operator form and applying the two requirements outlined in having a hybrid quantum code we arrive at a condition similar to the regular Knill--Laflamme condition but with $\alpha$ contained in the summations. Finally, if we consider the OAQEC model for the case where the entire space is correctable, namely that the noisy subspace is $1$. Then Eq.~\eqref{eqn:OAQECecc} simplifies to
\begin{equation}
P E_a^{\dagger}E_b P = \sum_k^M  \alpha_{kl}^{(\nu)} P_{Ak},
\end{equation}
which is equivalent to Eq.~\eqref{eqn:hybridrewritten}

\section{Hybrid Hamming bound} \label{sec:Hamming}

An important question in the discussion of hybrid codes is when hybrid codes will provide an advantage over codes which transmit quantum and classical information separately. Constructing the Hamming bound \cite{hamming1950error} for hybrid codes provides us with one means with which to compare the parameters of a hybrid and quantum code. The quantum Hamming bound applies to non-degenerate codes with the error set consisting of the Pauli matrices. For qubits, the bound can be stated as 
\begin{equation}
\displaystyle\sum_{j=0}^t \binom{n}{j}3^{j}2^{k} \leq 2^{n},
\end{equation}
where $t= \frac{d-1}{2}$. The bound can be reconstructed for hybrid codes. The quantum Hamming bound is essentially a packing argument. The argument is that the total space available to the system of qubits must be greater than the total space the errors can map codewords to plus the amount of space taken by the codewords themselves.

A code with $n$ physical qubits will have $2^n$ orthogonal subspaces available. Some of this space will be used by the logical codewords themselves, and the rest can be used by the space that errors map these codewords to. A code with $k$ qubits will have $2^k$ codewords. If the code can correct up to $j$ errors then there are $\binom{n}{j}$ sets of locations where an error can occur. At each location any of the three possible Pauli errors can occur, giving $3^j$ possible errors for each set of locations. These errors can occur on any of the $2^k$ codewords. This gives a total of $\sum_{j=1}^t \binom{n}{j}3^{j}2^{k}$ possible errors. Thus we arrive at the quantum Hamming bound,
\begin{align}
\sum_{j=1}^t \binom{n}{j}3^{j}2^{k} + 2^{k} &\leq 2^{n},\\
\sum_{j=0}^t \binom{n}{j}3^{j}2^{k} &\leq 2^{n}.
\end{align}
A hybrid code with $n$ physical qubits will also have a total of $2^n$ orthogonal subspaces available. For a hybrid code with $M$ codes, there will be $M2^k$  logical codewords. Since each quantum code making up the hybrid code must correct the same error set, then the number of locations an error can occur and the number of possible errors does not change for the non-degenerate case. The number of codewords this error can occur on has changed though to $M2^k$, thus the total number of errors which can occur is $\sum_{j=1}^t \binom{n}{j}3^{j}M2^{k}$. Therefore the quantum Hamming bound for hybrid codes  is given by:
\begin{align}
\displaystyle\sum_{j=1}^t \binom{n}{j}3^{j}M2^{k} + M2^{k} \leq 2^{n} \\
M\displaystyle\sum_{j=0}^t \binom{n}{j}3^{j}2^{k} \leq 2^{n} 
\end{align}
This bound highlights another important result of hybrid codes. A non-degenerate hybrid code can not provide an advantage over an equivalent quantum code. Therefore, degeneracy will be necessary for constructing non-trivial hybrid codes. We construct an example of a degenerate hybrid code. The code we construct has parameters $[[4,1:1,2]]$. This code can detect the error set $E = \{X_i, Y_i, Z_i, Z_1Z_2, Z_3Z_4 \} \quad \forall i = 1,2,3,4$, or equivalent in can correct the given error set, given the location of errors are known. This hybrid code has codewords
\begin{align}
\ket{\bar{0}}_0 = \ket{0000} + \ket{1111}, \quad \ket{\bar{1}}_0 = \ket{0011} - \ket{1100},   \label{eq:encoding1} \\
\ket{\bar{0}}_1 = \ket{0101} + \ket{1010}, \quad \ket{\bar{1}}_1 = \ket{1001} - \ket{0110}. \label{eq:encoding2}
\end{align}

\section{Outlook} \label{sec:conclusion}

We demonstrated that the two perspectives on hybrid codes are equivalent and that the coding theory characterization is a specific case of the OAQEC model. Furthermore, we generalized the quantum Hamming bound for hybrid error-correcting codes. In doing so, we found that a non-degenerate hybrid code can not provide an advantage over an equivalent quantum code. We then designed a four-qubit non-degenerate hybrid code. This code transmits one qubit and one classical bit and detects any single Pauli error.


Our results present natural opportunities for future work. First, there have been no physical implementations of hybrid codes in the literature. The code we constructed contained few enough qubits to be readily achieved on various quantum hardware available in academic labs. For instance, similar quantum states as those presented in Eq.~\eqref{eq:encoding1} and Eq.~\eqref{eq:encoding2} have been prepared using nuclear magnetic resonance spectrometers in, for example, tests of quantum error correction~\cite{cory1998experimental,leung1999experimental,knill2001benchmarking} and quantum foundations~\cite{majidy2019exploration,majidy2019violation,majidy2021detecting}. Second, developing the hybrid forms of other quantum bounds, particularly the quantum singleton bound and the quantum Gilbert--Varshamov bound~\cite{gilbert1952comparison,varshamov1957estimate}, is important for understanding the limitations of hybrid code. Finally, OAQEC has recently been used for various applications in the study of black holes~\cite{penington2020entanglement, kibe2022holographic,hayden2019learning,kim2020ghost}. Using the connection established in this work, it may be possible to cast a coding theory perspective on these black-hole physics results and benefit from the tools therein. 

\onecolumngrid
\bibliography{apssamp.bib}

\begin{thebibliography}{33}%
\makeatletter
\providecommand \@ifxundefined [1]{%
 \@ifx{#1\undefined}
}%
\providecommand \@ifnum [1]{%
 \ifnum #1\expandafter \@firstoftwo
 \else \expandafter \@secondoftwo
 \fi
}%
\providecommand \@ifx [1]{%
 \ifx #1\expandafter \@firstoftwo
 \else \expandafter \@secondoftwo
 \fi
}%
\providecommand \natexlab [1]{#1}%
\providecommand \enquote  [1]{``#1''}%
\providecommand \bibnamefont  [1]{#1}%
\providecommand \bibfnamefont [1]{#1}%
\providecommand \citenamefont [1]{#1}%
\providecommand \href@noop [0]{\@secondoftwo}%
\providecommand \href [0]{\begingroup \@sanitize@url \@href}%
\providecommand \@href[1]{\@@startlink{#1}\@@href}%
\providecommand \@@href[1]{\endgroup#1\@@endlink}%
\providecommand \@sanitize@url [0]{\catcode `\\12\catcode `\$12\catcode
  `\&12\catcode `\#12\catcode `\^12\catcode `\_12\catcode `\%12\relax}%
\providecommand \@@startlink[1]{}%
\providecommand \@@endlink[0]{}%
\providecommand \url  [0]{\begingroup\@sanitize@url \@url }%
\providecommand \@url [1]{\endgroup\@href {#1}{\urlprefix }}%
\providecommand \urlprefix  [0]{URL }%
\providecommand \Eprint [0]{\href }%
\providecommand \doibase [0]{https://doi.org/}%
\providecommand \selectlanguage [0]{\@gobble}%
\providecommand \bibinfo  [0]{\@secondoftwo}%
\providecommand \bibfield  [0]{\@secondoftwo}%
\providecommand \translation [1]{[#1]}%
\providecommand \BibitemOpen [0]{}%
\providecommand \bibitemStop [0]{}%
\providecommand \bibitemNoStop [0]{.\EOS\space}%
\providecommand \EOS [0]{\spacefactor3000\relax}%
\providecommand \BibitemShut  [1]{\csname bibitem#1\endcsname}%
\let\auto@bib@innerbib\@empty
\bibitem [{\citenamefont {Devetak}\ and\ \citenamefont
  {Shor}(2005)}]{devetak2005capacity}%
  \BibitemOpen
  \bibfield  {author} {\bibinfo {author} {\bibfnamefont {I.}~\bibnamefont
  {Devetak}}\ and\ \bibinfo {author} {\bibfnamefont {P.~W.}\ \bibnamefont
  {Shor}},\ }\bibfield  {title} {\bibinfo {title} {The capacity of a quantum
  channel for simultaneous transmission of classical and quantum information},\
  }\href@noop {} {\bibfield  {journal} {\bibinfo  {journal} {Communications in
  Mathematical Physics}\ }\textbf {\bibinfo {volume} {256}},\ \bibinfo {pages}
  {287} (\bibinfo {year} {2005})}\BibitemShut {NoStop}%
\bibitem [{\citenamefont {Hsieh}\ and\ \citenamefont
  {Wilde}(2010{\natexlab{a}})}]{hsieh2010entanglement}%
  \BibitemOpen
  \bibfield  {author} {\bibinfo {author} {\bibfnamefont {M.-H.}\ \bibnamefont
  {Hsieh}}\ and\ \bibinfo {author} {\bibfnamefont {M.~M.}\ \bibnamefont
  {Wilde}},\ }\bibfield  {title} {\bibinfo {title} {Entanglement-assisted
  communication of classical and quantum information},\ }\href@noop {}
  {\bibfield  {journal} {\bibinfo  {journal} {IEEE Transactions on Information
  Theory}\ }\textbf {\bibinfo {volume} {56}},\ \bibinfo {pages} {4682}
  (\bibinfo {year} {2010}{\natexlab{a}})}\BibitemShut {NoStop}%
\bibitem [{\citenamefont {Hsieh}\ and\ \citenamefont
  {Wilde}(2010{\natexlab{b}})}]{hsieh2010trading}%
  \BibitemOpen
  \bibfield  {author} {\bibinfo {author} {\bibfnamefont {M.-H.}\ \bibnamefont
  {Hsieh}}\ and\ \bibinfo {author} {\bibfnamefont {M.~M.}\ \bibnamefont
  {Wilde}},\ }\bibfield  {title} {\bibinfo {title} {Trading classical
  communication, quantum communication, and entanglement in quantum shannon
  theory},\ }\href@noop {} {\bibfield  {journal} {\bibinfo  {journal} {IEEE
  Transactions on Information Theory}\ }\textbf {\bibinfo {volume} {56}},\
  \bibinfo {pages} {4705} (\bibinfo {year} {2010}{\natexlab{b}})}\BibitemShut
  {NoStop}%
\bibitem [{\citenamefont {Kremsky}\ \emph {et~al.}(2008)\citenamefont
  {Kremsky}, \citenamefont {Hsieh},\ and\ \citenamefont
  {Brun}}]{kremsky2008classical}%
  \BibitemOpen
  \bibfield  {author} {\bibinfo {author} {\bibfnamefont {I.}~\bibnamefont
  {Kremsky}}, \bibinfo {author} {\bibfnamefont {M.-H.}\ \bibnamefont {Hsieh}},\
  and\ \bibinfo {author} {\bibfnamefont {T.~A.}\ \bibnamefont {Brun}},\
  }\bibfield  {title} {\bibinfo {title} {Classical enhancement of
  quantum-error-correcting codes},\ }\href@noop {} {\bibfield  {journal}
  {\bibinfo  {journal} {Physical Review A}\ }\textbf {\bibinfo {volume} {78}},\
  \bibinfo {pages} {012341} (\bibinfo {year} {2008})}\BibitemShut {NoStop}%
\bibitem [{\citenamefont {Nemec}(2022)}]{nemec2022hybrid}%
  \BibitemOpen
  \bibfield  {author} {\bibinfo {author} {\bibfnamefont {A.~S.}\ \bibnamefont
  {Nemec}},\ }\emph {\bibinfo {title} {Hybrid and Nonadditive Quantum Codes}},\
  \href@noop {} {Ph.D. thesis} (\bibinfo {year} {2022})\BibitemShut {NoStop}%
\bibitem [{\citenamefont {Nemec}\ and\ \citenamefont
  {Klappenecker}(2018)}]{nemec2018hybrid}%
  \BibitemOpen
  \bibfield  {author} {\bibinfo {author} {\bibfnamefont {A.}~\bibnamefont
  {Nemec}}\ and\ \bibinfo {author} {\bibfnamefont {A.}~\bibnamefont
  {Klappenecker}},\ }\bibfield  {title} {\bibinfo {title} {Hybrid codes},\ }in\
  \href@noop {} {\emph {\bibinfo {booktitle} {2018 IEEE International Symposium
  on Information Theory (ISIT)}}}\ (\bibinfo {organization} {IEEE},\ \bibinfo
  {year} {2018})\ pp.\ \bibinfo {pages} {796--800}\BibitemShut {NoStop}%
\bibitem [{\citenamefont {Nemec}\ and\ \citenamefont
  {Klappenecker}(2020)}]{nemec2020nonbinary}%
  \BibitemOpen
  \bibfield  {author} {\bibinfo {author} {\bibfnamefont {A.}~\bibnamefont
  {Nemec}}\ and\ \bibinfo {author} {\bibfnamefont {A.}~\bibnamefont
  {Klappenecker}},\ }\bibfield  {title} {\bibinfo {title} {Nonbinary
  error-detecting hybrid codes},\ }\href@noop {} {\bibfield  {journal}
  {\bibinfo  {journal} {American Journal of Science \& Engineering}\ }\textbf
  {\bibinfo {volume} {1}},\ \bibinfo {pages} {1} (\bibinfo {year}
  {2020})}\BibitemShut {NoStop}%
\bibitem [{\citenamefont {Nemec}\ and\ \citenamefont
  {Klappenecker}(2021)}]{nemec2021infinite}%
  \BibitemOpen
  \bibfield  {author} {\bibinfo {author} {\bibfnamefont {A.}~\bibnamefont
  {Nemec}}\ and\ \bibinfo {author} {\bibfnamefont {A.}~\bibnamefont
  {Klappenecker}},\ }\bibfield  {title} {\bibinfo {title} {Infinite families of
  quantum-classical hybrid codes},\ }\href@noop {} {\bibfield  {journal}
  {\bibinfo  {journal} {IEEE Transactions on Information Theory}\ }\textbf
  {\bibinfo {volume} {67}},\ \bibinfo {pages} {2847} (\bibinfo {year}
  {2021})}\BibitemShut {NoStop}%
\bibitem [{\citenamefont {Grassl}\ \emph {et~al.}(2017)\citenamefont {Grassl},
  \citenamefont {Lu},\ and\ \citenamefont {Zeng}}]{grassl2017codes}%
  \BibitemOpen
  \bibfield  {author} {\bibinfo {author} {\bibfnamefont {M.}~\bibnamefont
  {Grassl}}, \bibinfo {author} {\bibfnamefont {S.}~\bibnamefont {Lu}},\ and\
  \bibinfo {author} {\bibfnamefont {B.}~\bibnamefont {Zeng}},\ }\bibfield
  {title} {\bibinfo {title} {Codes for simultaneous transmission of quantum and
  classical information},\ }in\ \href@noop {} {\emph {\bibinfo {booktitle}
  {Information Theory (ISIT), 2017 IEEE International Symposium on}}}\
  (\bibinfo {organization} {IEEE},\ \bibinfo {year} {2017})\ pp.\ \bibinfo
  {pages} {1718--1722}\BibitemShut {NoStop}%
\bibitem [{\citenamefont {B{\'e}ny}\ \emph {et~al.}(2007)\citenamefont
  {B{\'e}ny}, \citenamefont {Kempf},\ and\ \citenamefont
  {Kribs}}]{beny2007quantum}%
  \BibitemOpen
  \bibfield  {author} {\bibinfo {author} {\bibfnamefont {C.}~\bibnamefont
  {B{\'e}ny}}, \bibinfo {author} {\bibfnamefont {A.}~\bibnamefont {Kempf}},\
  and\ \bibinfo {author} {\bibfnamefont {D.~W.}\ \bibnamefont {Kribs}},\
  }\bibfield  {title} {\bibinfo {title} {Quantum error correction of
  observables},\ }\href@noop {} {\bibfield  {journal} {\bibinfo  {journal}
  {Physical Review A}\ }\textbf {\bibinfo {volume} {76}},\ \bibinfo {pages}
  {042303} (\bibinfo {year} {2007})}\BibitemShut {NoStop}%
\bibitem [{\citenamefont {Gottesman}(1996)}]{gottesman1996class}%
  \BibitemOpen
  \bibfield  {author} {\bibinfo {author} {\bibfnamefont {D.}~\bibnamefont
  {Gottesman}},\ }\bibfield  {title} {\bibinfo {title} {Class of quantum
  error-correcting codes saturating the quantum hamming bound},\ }\href@noop {}
  {\bibfield  {journal} {\bibinfo  {journal} {Physical Review A}\ }\textbf
  {\bibinfo {volume} {54}},\ \bibinfo {pages} {1862} (\bibinfo {year}
  {1996})}\BibitemShut {NoStop}%
\bibitem [{\citenamefont {Knill}\ and\ \citenamefont
  {Laflamme}(1997)}]{knill1997theory}%
  \BibitemOpen
  \bibfield  {author} {\bibinfo {author} {\bibfnamefont {E.}~\bibnamefont
  {Knill}}\ and\ \bibinfo {author} {\bibfnamefont {R.}~\bibnamefont
  {Laflamme}},\ }\bibfield  {title} {\bibinfo {title} {Theory of quantum
  error-correcting codes},\ }\href@noop {} {\bibfield  {journal} {\bibinfo
  {journal} {Physical Review A}\ }\textbf {\bibinfo {volume} {55}},\ \bibinfo
  {pages} {900} (\bibinfo {year} {1997})}\BibitemShut {NoStop}%
\bibitem [{\citenamefont {Kribs}\ \emph {et~al.}(2005)\citenamefont {Kribs},
  \citenamefont {Laflamme},\ and\ \citenamefont {Poulin}}]{kribs2005unified}%
  \BibitemOpen
  \bibfield  {author} {\bibinfo {author} {\bibfnamefont {D.}~\bibnamefont
  {Kribs}}, \bibinfo {author} {\bibfnamefont {R.}~\bibnamefont {Laflamme}},\
  and\ \bibinfo {author} {\bibfnamefont {D.}~\bibnamefont {Poulin}},\
  }\bibfield  {title} {\bibinfo {title} {Unified and generalized approach to
  quantum error correction},\ }\href@noop {} {\bibfield  {journal} {\bibinfo
  {journal} {Physical review letters}\ }\textbf {\bibinfo {volume} {94}},\
  \bibinfo {pages} {180501} (\bibinfo {year} {2005})}\BibitemShut {NoStop}%
\bibitem [{\citenamefont {Shor}(1995)}]{shor1995scheme}%
  \BibitemOpen
  \bibfield  {author} {\bibinfo {author} {\bibfnamefont {P.~W.}\ \bibnamefont
  {Shor}},\ }\bibfield  {title} {\bibinfo {title} {Scheme for reducing
  decoherence in quantum computer memory},\ }\href@noop {} {\bibfield
  {journal} {\bibinfo  {journal} {Physical review A}\ }\textbf {\bibinfo
  {volume} {52}},\ \bibinfo {pages} {R2493} (\bibinfo {year}
  {1995})}\BibitemShut {NoStop}%
\bibitem [{\citenamefont {Steane}(1996)}]{steane1996simple}%
  \BibitemOpen
  \bibfield  {author} {\bibinfo {author} {\bibfnamefont {A.~M.}\ \bibnamefont
  {Steane}},\ }\bibfield  {title} {\bibinfo {title} {Simple quantum
  error-correcting codes},\ }\href@noop {} {\bibfield  {journal} {\bibinfo
  {journal} {Physical Review A}\ }\textbf {\bibinfo {volume} {54}},\ \bibinfo
  {pages} {4741} (\bibinfo {year} {1996})}\BibitemShut {NoStop}%
\bibitem [{\citenamefont {Bennett}\ \emph {et~al.}(1996)\citenamefont
  {Bennett}, \citenamefont {DiVincenzo}, \citenamefont {Smolin},\ and\
  \citenamefont {Wootters}}]{bennett1996ch}%
  \BibitemOpen
  \bibfield  {author} {\bibinfo {author} {\bibfnamefont {C.~H.}\ \bibnamefont
  {Bennett}}, \bibinfo {author} {\bibfnamefont {D.~P.}\ \bibnamefont
  {DiVincenzo}}, \bibinfo {author} {\bibfnamefont {J.~A.}\ \bibnamefont
  {Smolin}},\ and\ \bibinfo {author} {\bibfnamefont {W.~K.}\ \bibnamefont
  {Wootters}},\ }\bibfield  {title} {\bibinfo {title} {Mixed-state entanglement
  and quantum error correction},\ }\href@noop {} {\bibfield  {journal}
  {\bibinfo  {journal} {Physical Review A}\ }\textbf {\bibinfo {volume} {54}},\
  \bibinfo {pages} {3824} (\bibinfo {year} {1996})}\BibitemShut {NoStop}%
\bibitem [{\citenamefont {Palma}\ \emph {et~al.}(1996)\citenamefont {Palma},
  \citenamefont {Suominen},\ and\ \citenamefont {Ekert}}]{palma1996quantum}%
  \BibitemOpen
  \bibfield  {author} {\bibinfo {author} {\bibfnamefont {G.~M.}\ \bibnamefont
  {Palma}}, \bibinfo {author} {\bibfnamefont {K.-A.}\ \bibnamefont
  {Suominen}},\ and\ \bibinfo {author} {\bibfnamefont {A.}~\bibnamefont
  {Ekert}},\ }\bibfield  {title} {\bibinfo {title} {Quantum computers and
  dissipation},\ }\href@noop {} {\bibfield  {journal} {\bibinfo  {journal}
  {Proceedings of the Royal Society of London. Series A: Mathematical, Physical
  and Engineering Sciences}\ }\textbf {\bibinfo {volume} {452}},\ \bibinfo
  {pages} {567} (\bibinfo {year} {1996})}\BibitemShut {NoStop}%
\bibitem [{\citenamefont {Duan}\ and\ \citenamefont
  {Guo}(1997)}]{duan1997preserving}%
  \BibitemOpen
  \bibfield  {author} {\bibinfo {author} {\bibfnamefont {L.-M.}\ \bibnamefont
  {Duan}}\ and\ \bibinfo {author} {\bibfnamefont {G.-C.}\ \bibnamefont {Guo}},\
  }\bibfield  {title} {\bibinfo {title} {Preserving coherence in quantum
  computation by pairing quantum bits},\ }\href@noop {} {\bibfield  {journal}
  {\bibinfo  {journal} {Physical Review Letters}\ }\textbf {\bibinfo {volume}
  {79}},\ \bibinfo {pages} {1953} (\bibinfo {year} {1997})}\BibitemShut
  {NoStop}%
\bibitem [{\citenamefont {Zanardi}\ and\ \citenamefont
  {Rasetti}(1997)}]{zanardi1997noiseless}%
  \BibitemOpen
  \bibfield  {author} {\bibinfo {author} {\bibfnamefont {P.}~\bibnamefont
  {Zanardi}}\ and\ \bibinfo {author} {\bibfnamefont {M.}~\bibnamefont
  {Rasetti}},\ }\bibfield  {title} {\bibinfo {title} {Noiseless quantum
  codes},\ }\href@noop {} {\bibfield  {journal} {\bibinfo  {journal} {Physical
  Review Letters}\ }\textbf {\bibinfo {volume} {79}},\ \bibinfo {pages} {3306}
  (\bibinfo {year} {1997})}\BibitemShut {NoStop}%
\bibitem [{\citenamefont {Lidar}\ \emph {et~al.}(1998)\citenamefont {Lidar},
  \citenamefont {Chuang},\ and\ \citenamefont {Whaley}}]{lidar1998decoherence}%
  \BibitemOpen
  \bibfield  {author} {\bibinfo {author} {\bibfnamefont {D.~A.}\ \bibnamefont
  {Lidar}}, \bibinfo {author} {\bibfnamefont {I.~L.}\ \bibnamefont {Chuang}},\
  and\ \bibinfo {author} {\bibfnamefont {K.~B.}\ \bibnamefont {Whaley}},\
  }\bibfield  {title} {\bibinfo {title} {Decoherence-free subspaces for quantum
  computation},\ }\href@noop {} {\bibfield  {journal} {\bibinfo  {journal}
  {Physical Review Letters}\ }\textbf {\bibinfo {volume} {81}},\ \bibinfo
  {pages} {2594} (\bibinfo {year} {1998})}\BibitemShut {NoStop}%
\bibitem [{\citenamefont {Hamming}(1950)}]{hamming1950error}%
  \BibitemOpen
  \bibfield  {author} {\bibinfo {author} {\bibfnamefont {R.~W.}\ \bibnamefont
  {Hamming}},\ }\bibfield  {title} {\bibinfo {title} {Error detecting and error
  correcting codes},\ }\href@noop {} {\bibfield  {journal} {\bibinfo  {journal}
  {The Bell system technical journal}\ }\textbf {\bibinfo {volume} {29}},\
  \bibinfo {pages} {147} (\bibinfo {year} {1950})}\BibitemShut {NoStop}%
\bibitem [{\citenamefont {Cory}\ \emph {et~al.}(1998)\citenamefont {Cory},
  \citenamefont {Price}, \citenamefont {Maas}, \citenamefont {Knill},
  \citenamefont {Laflamme}, \citenamefont {Zurek}, \citenamefont {Havel},\ and\
  \citenamefont {Somaroo}}]{cory1998experimental}%
  \BibitemOpen
  \bibfield  {author} {\bibinfo {author} {\bibfnamefont {D.~G.}\ \bibnamefont
  {Cory}}, \bibinfo {author} {\bibfnamefont {M.}~\bibnamefont {Price}},
  \bibinfo {author} {\bibfnamefont {W.}~\bibnamefont {Maas}}, \bibinfo {author}
  {\bibfnamefont {E.}~\bibnamefont {Knill}}, \bibinfo {author} {\bibfnamefont
  {R.}~\bibnamefont {Laflamme}}, \bibinfo {author} {\bibfnamefont {W.~H.}\
  \bibnamefont {Zurek}}, \bibinfo {author} {\bibfnamefont {T.~F.}\ \bibnamefont
  {Havel}},\ and\ \bibinfo {author} {\bibfnamefont {S.~S.}\ \bibnamefont
  {Somaroo}},\ }\bibfield  {title} {\bibinfo {title} {Experimental quantum
  error correction},\ }\href@noop {} {\bibfield  {journal} {\bibinfo  {journal}
  {Physical Review Letters}\ }\textbf {\bibinfo {volume} {81}},\ \bibinfo
  {pages} {2152} (\bibinfo {year} {1998})}\BibitemShut {NoStop}%
\bibitem [{\citenamefont {Leung}\ \emph {et~al.}(1999)\citenamefont {Leung},
  \citenamefont {Vandersypen}, \citenamefont {Zhou}, \citenamefont {Sherwood},
  \citenamefont {Yannoni}, \citenamefont {Kubinec},\ and\ \citenamefont
  {Chuang}}]{leung1999experimental}%
  \BibitemOpen
  \bibfield  {author} {\bibinfo {author} {\bibfnamefont {D.}~\bibnamefont
  {Leung}}, \bibinfo {author} {\bibfnamefont {L.}~\bibnamefont {Vandersypen}},
  \bibinfo {author} {\bibfnamefont {X.}~\bibnamefont {Zhou}}, \bibinfo {author}
  {\bibfnamefont {M.}~\bibnamefont {Sherwood}}, \bibinfo {author}
  {\bibfnamefont {C.}~\bibnamefont {Yannoni}}, \bibinfo {author} {\bibfnamefont
  {M.}~\bibnamefont {Kubinec}},\ and\ \bibinfo {author} {\bibfnamefont
  {I.}~\bibnamefont {Chuang}},\ }\bibfield  {title} {\bibinfo {title}
  {Experimental realization of a two-bit phase damping quantum code},\
  }\href@noop {} {\bibfield  {journal} {\bibinfo  {journal} {Physical Review
  A}\ }\textbf {\bibinfo {volume} {60}},\ \bibinfo {pages} {1924} (\bibinfo
  {year} {1999})}\BibitemShut {NoStop}%
\bibitem [{\citenamefont {Knill}\ \emph {et~al.}(2001)\citenamefont {Knill},
  \citenamefont {Laflamme}, \citenamefont {Martinez},\ and\ \citenamefont
  {Negrevergne}}]{knill2001benchmarking}%
  \BibitemOpen
  \bibfield  {author} {\bibinfo {author} {\bibfnamefont {E.}~\bibnamefont
  {Knill}}, \bibinfo {author} {\bibfnamefont {R.}~\bibnamefont {Laflamme}},
  \bibinfo {author} {\bibfnamefont {R.}~\bibnamefont {Martinez}},\ and\
  \bibinfo {author} {\bibfnamefont {C.}~\bibnamefont {Negrevergne}},\
  }\bibfield  {title} {\bibinfo {title} {Benchmarking quantum computers: the
  five-qubit error correcting code},\ }\href@noop {} {\bibfield  {journal}
  {\bibinfo  {journal} {Physical Review Letters}\ }\textbf {\bibinfo {volume}
  {86}},\ \bibinfo {pages} {5811} (\bibinfo {year} {2001})}\BibitemShut
  {NoStop}%
\bibitem [{\citenamefont {Majidy}\ \emph {et~al.}(2019)\citenamefont {Majidy},
  \citenamefont {Katiyar}, \citenamefont {Anikeeva}, \citenamefont
  {Halliwell},\ and\ \citenamefont {Laflamme}}]{majidy2019exploration}%
  \BibitemOpen
  \bibfield  {author} {\bibinfo {author} {\bibfnamefont {S.-S.}\ \bibnamefont
  {Majidy}}, \bibinfo {author} {\bibfnamefont {H.}~\bibnamefont {Katiyar}},
  \bibinfo {author} {\bibfnamefont {G.}~\bibnamefont {Anikeeva}}, \bibinfo
  {author} {\bibfnamefont {J.}~\bibnamefont {Halliwell}},\ and\ \bibinfo
  {author} {\bibfnamefont {R.}~\bibnamefont {Laflamme}},\ }\bibfield  {title}
  {\bibinfo {title} {Exploration of an augmented set of leggett-garg
  inequalities using a noninvasive continuous-in-time velocity measurement},\
  }\href@noop {} {\bibfield  {journal} {\bibinfo  {journal} {Physical Review
  A}\ }\textbf {\bibinfo {volume} {100}},\ \bibinfo {pages} {042325} (\bibinfo
  {year} {2019})}\BibitemShut {NoStop}%
\bibitem [{\citenamefont {Majidy}(2019)}]{majidy2019violation}%
  \BibitemOpen
  \bibfield  {author} {\bibinfo {author} {\bibfnamefont {S.-S.}\ \bibnamefont
  {Majidy}},\ }\emph {\bibinfo {title} {Violation of an augmented set of
  Leggett-Garg inequalities and the implementation of a continuous in time
  velocity measurement}},\ \href@noop {} {Master's thesis},\ \bibinfo  {school}
  {University of Waterloo} (\bibinfo {year} {2019})\BibitemShut {NoStop}%
\bibitem [{\citenamefont {Majidy}\ \emph {et~al.}(2021)\citenamefont {Majidy},
  \citenamefont {Halliwell},\ and\ \citenamefont
  {Laflamme}}]{majidy2021detecting}%
  \BibitemOpen
  \bibfield  {author} {\bibinfo {author} {\bibfnamefont {S.}~\bibnamefont
  {Majidy}}, \bibinfo {author} {\bibfnamefont {J.~J.}\ \bibnamefont
  {Halliwell}},\ and\ \bibinfo {author} {\bibfnamefont {R.}~\bibnamefont
  {Laflamme}},\ }\bibfield  {title} {\bibinfo {title} {Detecting violations of
  macrorealism when the original leggett-garg inequalities are satisfied},\
  }\href@noop {} {\bibfield  {journal} {\bibinfo  {journal} {Physical Review
  A}\ }\textbf {\bibinfo {volume} {103}},\ \bibinfo {pages} {062212} (\bibinfo
  {year} {2021})}\BibitemShut {NoStop}%
\bibitem [{\citenamefont {Gilbert}(1952)}]{gilbert1952comparison}%
  \BibitemOpen
  \bibfield  {author} {\bibinfo {author} {\bibfnamefont {E.~N.}\ \bibnamefont
  {Gilbert}},\ }\bibfield  {title} {\bibinfo {title} {A comparison of
  signalling alphabets},\ }\href@noop {} {\bibfield  {journal} {\bibinfo
  {journal} {The Bell system technical journal}\ }\textbf {\bibinfo {volume}
  {31}},\ \bibinfo {pages} {504} (\bibinfo {year} {1952})}\BibitemShut
  {NoStop}%
\bibitem [{\citenamefont {Varshamov}(1957)}]{varshamov1957estimate}%
  \BibitemOpen
  \bibfield  {author} {\bibinfo {author} {\bibfnamefont {R.~R.}\ \bibnamefont
  {Varshamov}},\ }\bibfield  {title} {\bibinfo {title} {Estimate of the number
  of signals in error correcting codes},\ }\href@noop {} {\bibfield  {journal}
  {\bibinfo  {journal} {Docklady Akad. Nauk, SSSR}\ }\textbf {\bibinfo {volume}
  {117}},\ \bibinfo {pages} {739} (\bibinfo {year} {1957})}\BibitemShut
  {NoStop}%
\bibitem [{\citenamefont {Penington}(2020)}]{penington2020entanglement}%
  \BibitemOpen
  \bibfield  {author} {\bibinfo {author} {\bibfnamefont {G.}~\bibnamefont
  {Penington}},\ }\bibfield  {title} {\bibinfo {title} {Entanglement wedge
  reconstruction and the information paradox},\ }\href@noop {} {\bibfield
  {journal} {\bibinfo  {journal} {Journal of High Energy Physics}\ }\textbf
  {\bibinfo {volume} {2020}},\ \bibinfo {pages} {1} (\bibinfo {year}
  {2020})}\BibitemShut {NoStop}%
\bibitem [{\citenamefont {Kibe}\ \emph {et~al.}(2022)\citenamefont {Kibe},
  \citenamefont {Mandayam},\ and\ \citenamefont
  {Mukhopadhyay}}]{kibe2022holographic}%
  \BibitemOpen
  \bibfield  {author} {\bibinfo {author} {\bibfnamefont {T.}~\bibnamefont
  {Kibe}}, \bibinfo {author} {\bibfnamefont {P.}~\bibnamefont {Mandayam}},\
  and\ \bibinfo {author} {\bibfnamefont {A.}~\bibnamefont {Mukhopadhyay}},\
  }\bibfield  {title} {\bibinfo {title} {Holographic spacetime, black holes and
  quantum error correcting codes: a review},\ }\href@noop {} {\bibfield
  {journal} {\bibinfo  {journal} {The European Physical Journal C}\ }\textbf
  {\bibinfo {volume} {82}},\ \bibinfo {pages} {463} (\bibinfo {year}
  {2022})}\BibitemShut {NoStop}%
\bibitem [{\citenamefont {Hayden}\ and\ \citenamefont
  {Penington}(2019)}]{hayden2019learning}%
  \BibitemOpen
  \bibfield  {author} {\bibinfo {author} {\bibfnamefont {P.}~\bibnamefont
  {Hayden}}\ and\ \bibinfo {author} {\bibfnamefont {G.}~\bibnamefont
  {Penington}},\ }\bibfield  {title} {\bibinfo {title} {Learning the alpha-bits
  of black holes},\ }\href@noop {} {\bibfield  {journal} {\bibinfo  {journal}
  {Journal of High Energy Physics}\ }\textbf {\bibinfo {volume} {2019}},\
  \bibinfo {pages} {1} (\bibinfo {year} {2019})}\BibitemShut {NoStop}%
\bibitem [{\citenamefont {Kim}\ \emph {et~al.}(2020)\citenamefont {Kim},
  \citenamefont {Tang},\ and\ \citenamefont {Preskill}}]{kim2020ghost}%
  \BibitemOpen
  \bibfield  {author} {\bibinfo {author} {\bibfnamefont {I.}~\bibnamefont
  {Kim}}, \bibinfo {author} {\bibfnamefont {E.}~\bibnamefont {Tang}},\ and\
  \bibinfo {author} {\bibfnamefont {J.}~\bibnamefont {Preskill}},\ }\bibfield
  {title} {\bibinfo {title} {The ghost in the radiation: Robust encodings of
  the black hole interior},\ }\href@noop {} {\bibfield  {journal} {\bibinfo
  {journal} {Journal of High Energy Physics}\ }\textbf {\bibinfo {volume}
  {2020}},\ \bibinfo {pages} {1} (\bibinfo {year} {2020})}\BibitemShut
  {NoStop}%
\end{thebibliography}%

\end{document}